\documentclass[a4paper]{jpconf}
\usepackage{graphicx}
\newcommand{\beq}{\begin{equation}}
\newcommand{\eeq}{\end{equation}}
\begin{document}
\title{Chiral Random Matrix Theory and Chiral Perturbation Theory}

\author{Poul H. Damgaard}

\address{The Niels Bohr International Academy and Discovery Center, 
The Niels Bohr Institute, Blegdamsvej 17, DK-2100 Copenhagen, Denmark}

\ead{phdamg@nbi.dk}

\begin{abstract}
Spontaneous breaking of chiral symmetry in QCD has traditionally 
been inferred indirectly through low-energy theorems and comparison
with experiments. Thanks to the understanding of an unexpected connection
between chiral Random Matrix Theory and chiral Perturbation Theory, 
the spontaneous breaking of chiral symmetry in QCD can now be shown
unequivocally from first principles and lattice simulations. In these
lectures I give an introduction to the subject, starting with
an elementary discussion of spontaneous breaking of global symmetries. 
\end{abstract}

\section{Introduction}

Many tend to think of spontaneous breaking of global symmetries in quantum
field theory as something rather simple, almost trivial. To illustrate 
that this is certainly not so, let us first briefly review the standard 
textbook treatment of this phenomenon. Typically, one considers a
field theory of a real scalar field 
with a potential that, appropriately for this
school, can be dubbed a ``sombrero potential'',
\beq
V(\phi) ~=~ -\frac{1}{2}m^2\phi(x)^2 + \frac{\lambda}{4!}\phi(x)^4 ~.
\eeq  
{}From a perspective of classical physics this looks unstable.
Certainly a constant field configuration of $\phi(x) = 0$ cannot
possibly be a good starting point for a perturbative expansion.
Instead, one considers the two minima of the potential at
\beq
\phi ~=~ \pm v ~=~ \pm \sqrt{\frac{6}{\lambda}}m
\eeq
and says that a consistent treatment of such a field theory
must be based on one vacuum that is {\em either} at $\phi = v$
or at $\phi = -v$, which one being undecidable.
Here a comparison is often made to a classical picture, like a pencil
standing exactly on its tip: such a system is rotationally symmetric,
but any small perturbation will evidently make the pencil fall in
the direction of the perturbation. Spontaneous symmetry breaking!
Indeed, the pencil will now lie flat on the table, and it will
have picked just one random direction, the one that was induced by
the perturbation, a perturbation we can make as small as we like. What
about the potential energy of the pencil? It was first converted to
kinetic energy, and then, on impact with the table, to heat. Such
a system is dissipative: If it were not, the pencil would simply
bounce back up. There may issues such as accuracy of initial conditions
and so on, but this last point illustrates the difficulty with
a simple picture of spontaneous symmetry breaking. 

To investigate this a little more closely, let us proceed to
quantum mechanics and glue two 
harmonic oscillator potentials together, separated by a distance
$2a$\footnote{This is also a standard textbook example,
see $e.g.$, ref. \cite{Merzbacher} for a nice and clear discussion.}.
The potential is thus
\beq
V(x) ~=~ \frac{1}{2}m\omega^2\left(|x| - a\right)^2
\eeq
The ground state of this quantum mechanical system must have no
nodes, and we can almost trivially identify the correct ground
state by drawing two Gaussian bumps centered at $x=\pm a$, 
\beq
\psi_{\pm}(x) ~\sim~ \exp\left[-\frac{1}{2}m\omega(x\mp a)^2\right]
\eeq
which are, individually, ground states of either of the two 
harmonic oscillators. If we smoothly join the two at $x=0$ so
that we construct, roughly, 
\beq 
\psi_0(x) ~\sim~ \psi_+(x) + \psi_-(x)
\eeq
this will be a very good approximation to the true ground state.
What about the 1st excited state? It should have one node, and again
we can easily guess its form. Suppose we glue the two wave 
functions $\psi_{\pm}(x)$ together, but with a twist: we flip
the sign of one of them to get
\beq
\psi_1(x) ~\sim~  \psi_+(x) - \psi_-(x)
\eeq
This state will have a {\em slightly} higher energy than $\psi_0(x)$,
but it is clear from this simple construction that $\psi_0(x)$ and
$\psi_1(x)$ are almost degenerate. The higher excited states can
be visualized similarly, by gluing two wave functions together around
$x=0$, more and more excited states, and alternatingly adding and
subtracting combinations.

Does such a system display spontaneous symmetry breaking? It is well
known that in this quantum mechanical case
one of the two apparently distinct vacua will never be selected. Although
the first excited state is almost degenerate with the ground state,
there will always be a finite energy difference $E_1-E_0$. If we 
prepare a state that initially is localized in only one of the two
wells, it will tunnel to the other side. The tunneling time
goes like $(E_1-E_0)^{-1}$. Only when the two minima are infinitely
far from each other does that tunneling time go to infinity. So our
naive picture of spontaneous symmetry breaking can clearly not
be valid for such a simple quantum mechanical system of just one
degree of freedom. One cannot contain the probability density 
on just one side of $x=0$.

These two examples, one classical and one quantum mechanical, illustrate
that the phenomenon of spontaneous symmetry breaking in quantum field
theory cannot be quite as simple as the intuition based on the
sombrero potential seems to suggest. Nevertheless, as is so often
the case in physics: the argument may not be quite right, but the answer
is correct. There is no spontaneous breaking of symmetry in quantum
mechanical systems of a finite number of degrees of freedom. But there
{\em can} be spontaneous breaking of symmetries in quantum field
theory. The distinction here comes precisely from the infinite
number of degrees of freedom, the infinite number of qauntum field theory
modes. Although each of them individually can prevent spontaneous
breaking of symmetry by tunneling, this mechanism can be blocked
when an infinite number of modes have to tunnel. Subtleties in the
arguments have been discussed on and off in the literature, see for
instance refs. \cite{Banks,Nieto}. For a more recent discussion, with
plenty of references to earlier literature, see ref. \cite{Perez}.

To make the confusion complete, let us also point out another
related phenomenon in quantum field theory: tunneling from
vacuum to another. We have just argued that in quantum field
theory the tunneling between degenerate vacua can be prohibited
due to the fact that an infinite number of degrees of freedom
have to simultaneously align. Yet, if we tilt the sombrero
potential slightly so that one vacuum really becomes energetically
favored it is also common knowledge that in quantum field
theory we {\em do} have tunneling to the true vacuum. How can this
be? It turns out to be another example of a superficial argument
that is basically correct nevetheless, and there is an endless
amount of literature on this subject. Suffice it to say here
that when there is a genuine non-degeneracy of vacua, one can
have bubble formation and a first order phase transition where
the true vacuum gobbles up the false vacuum. For a nice discussion
of the essential physics of this situation see, $e.g.$, ref.
\cite{Voloshin}. 

Of particular importance is the phenomenon of spontaneous breaking
of continuous global symmetries because it is linked to what is
known as Goldstone's Theorem. This theorem implies the appearance
of one massless mode associated with each broken generator of the
continuous group of symmetries. Again there are caveats, and there
are several. One is space-time dimensionality: a massless scalar
in two space-time dimensions is not well defined in an infinite
volume due to infrared singularities, and indeed there is no
Goldstone phenomenon in two space-time dimensions \cite{Coleman}.
In higher dimensions the theorem is incredibly strong, as it 
says that there is no mass gap, an otherwise forbiddingly difficult
question to tackle. The Goldstone Theorem is most concisely stated in the
operator language. Since the symmetry is continuous, let us consider the
associated conserved current operator. Ordinarily the vacuum $|0\rangle$
is assumed to
be annihilated by the corresponding conserved charge operator $Q$
(the vacuum is not charged under $Q$). If the vacuum is 
{\em} not annihilated by $Q$, the symmetry
is spontaneously broken. Remarkably, this implies the existence of
an associated massless mode. This is the content of Goldstone's Theorem.

In QCD, the spontaneous breaking of chiral symmetry is a profound
phenomenon that has been inferred indirectly over a long period
of both theoretical and experimental development, 
starting from the days when the fundamental
theory of strong interactions was not even known. In hindsight, it
is incredible that it could be gotten at without a correct understanding
of the underlying microscopic mechanism. Today we're much better off:
we know the fundamental Lagrangian of QCD and it is ``just'' a matter
of checking whether chiral symmetries are broken spontaneously or not.
In practice, this is not simple {\em at all}. In fact, the question
of showing spontaneous breaking of chiral symmetry in QCD and QCD-like
theories is a challenge at the level of showing confinement and the
existence of a mass gap in pure Yang-Mills theory. It is one of the
remarkable successes of the numerical non-perturbative approach to
QCD -- lattice QCD -- that this now has been shown beyond any doubt.
An essential ingredient in this comes from the new understanding
of chiral symmetry breaking through Random Matrix Theory. 

The purpose of these very elementary lectures was to acquaint students with 
the new
developments in our understanding of chiral symmetry breaking in QCD.
The amount of time was insufficient to give a complete review
and this is reflected in these notes which have focused on only
a few of the many interesting aspects.
That chiral Random Matrix Theory can say something {\em exact}
about a complicated quantum field theory such as QCD with light quarks
sounds like a wild idea, and it 
was initially met by a lot of skepticism in the lattice QCD community.
All of this skepticism turned out to be unwarranted. 
There is now complete understanding of how and why
chiral Random Matrix Theory represents an exact limit of the light
quark partition function of QCD in a specific finite-volume regime.
A precise mapping can be made between observables computed in QCD
in that regime, the standard chiral Lagrangian of light-quark QCD
and chiral Random Matrix Theory. This is a fantastic achievement,
a most surprising and deep relation between a quantum field theory
as complicated as QCD and a sequence of universal phenomena that
follow from the spontaneous breaking of chiral symmetry alone. One
example suffices to illustrate the enormous progress that has been
achieved. Let us consider the eigenvalues of the QCD Dirac operator $D$:
\beq
D\psi_n ~=~ \lambda_n\psi_n
\eeq
To make the discussing well-defined, let us consider this eigenvalue
problem in a theory with a finite ultraviolet cut-off $\Lambda$ and
a finite infrared cut-off $L$.
As will be discussed further down in these lectures, spontaneous
breaking of chiral symmetry requires that the {\em smallest} 
non-zero eigenvalues $\lambda_n$ accumulate towards zero at a rate
proportional to $1/L^4$ in four dimensions. This is thus a simple
test: do they or do they not? Before the developments described here, 
the strongest statement that could be made on basic principles 
was a proof that the $\lambda_n$'s accumulate {\em at least} as
fast as $1/L$ \cite{VW}. But this is a very weak condition. Even
in a free theory, eigenvalues accumulate as $1/L$. All the theorem says, then,
is that Dirac eigenvalues in QCD do not accumulate with a rate that is
{\em slower} than that of free fermions, $i.e.$, as if there were no gauge 
interactions whatsoever. Can one not do better?
With the advent of chiral Random Matrix Theory
this problem has been solved, and it is now known that in QCD with
two light flavors (the $u$ and $d$ quarks), Dirac operator eigenvalues
{\em do} accumulate towards the origin at a much faster rate that goes 
exactly as $1/L^4$. Not only that, the precise probability distributions
of single individual Dirac eigenvalues, $\lambda_i, i = 1, 2, 3,\ldots$
ordered according to their distance from the origin can be computed
exactly. These distributions, as well as all spectral correlation
functions of these smallest Dirac operator eigenvalues follow
universal scaling laws whose exact analytical forms are known.  

\section{Chiral symmetry in QCD}

If we look at the Lagrangian density of QCD with massless quarks,
we are struck by the fact that left-handed and right-handed
field decouple. If we had exactly $N_f$ massless
quarks in QCD, the global symmetry group would be $U(N_f)\times U(N_f)$.
Of this a subgroup is $U(1)$ baryon number, which remains unbroken
in QCD. The singlet chiral $U(1)$ symmetry is apparent only: it is
broken by the chiral anomaly, and is therefore not a symmetry of
the quantum theory. What remain are two independent flavor rotations
$SU(N_f)$ for the left and right handed quarks. If this symmetry
is broken as expected, $SU(N_f)_L\times SU(N_F)_R \to SU(N_f)$
due to the formation of a
flavor-independent chiral condensate $\langle\bar{\psi}\psi\rangle$,
we have $N_f^2-1$ broken generators and hence, by Goldstone's
Theorem, $N_f^2-1$ massless bosons. This is a profound statement!
How this slowly came to be realized before quarks had even been
introduced is a story in itself. One consequence of this is that
a low-energy representation of the QCD partition function had been
inferred long before QCD was considered. In the next subsection we
will give a lightning review of the moderne viewpoint on this.

\subsection{The Chiral Lagrangian}

Because of Goldstone's Theorem, the low-energy degrees of freedom
of massless QCD are those of the Nambu-Goldstone bosons. In reality,
even the $u$ and $d$ quarks are not exactly massless. But on the
QCD scale of $\sim$ 1 Gev they look nearly massless; their
masses are on the order a few MeV. At some point it must be possible
to treat the quark masses as small perturbations on top of a
theory of genuine Goldstone bosons. The chiral Lagrangian does 
exactly this. Consider first a truly massless $N_f$-flavor theory
where chiral symmetry is spontaneously broken as discussed above.
To describe the coset of chiral symmetry breaking it is convenient
to introduce an unusual non-linear field representation in terms
of group elements $U(x) = \exp[i\sqrt{2}\Phi(x)/F]$, where $F$ is
the pion decay constant and $\Phi(x) =\lambda^a\phi^{a}/\sqrt{2}$
represents the collection of $N_f^2-1$ pion fields.
A field theory based on this should have the following ingredients: 
(1) kinetic energy terms for the pions with canonical normalization,
(2) vanishing interactions at zero energy (as follows from 
Goldstone's Theorem) and (3) invariance under the chiral rotations. 
In the absence of any other principle
that could fix terms in the Lagrangian the only solution will be to
write down {\em all} terms that comply with the two conditions above.
This gives us an endless series of invariants:
\beq
{\cal L} = \frac{F^2}{4}Tr[\partial_{\mu}U^{\dagger}\partial^{\mu}U]
+ L_1({\rm Tr}[\partial_{\mu}U^{\dagger}\partial^{\mu}U])^2 + 
L_2{\rm Tr}[\partial_{\mu}U^{\dagger}\partial_{\nu}U]
{\rm Tr}[\partial^{\nu}U^{\dagger}\partial^{\mu}U] +
\ldots \label{Lchi}
\eeq
with an infinite series of couplings (we display only the first two,
$L_1$ and $L_2$ here). Values of all these couplings 
are left totally unspecified.
They must depend on details of the dynamics (which gauge group
is responsible for the symmetry breaking and so on). This is the
chiral Lagrangian.

This cannot be the full story because we also want to treat quark
masses, at least in an as yet not totally specified perturbation to
the above Lagrangian. There is a beautifully simple way to do this,
which goes under the name of the spurion technique. The chiral Lagrangian  
(\ref{Lchi}) is by construction invariant under global transformations
$U \to U_R U U_L^{\dagger}$. A mass term in the QCD Lagrangian,
\beq
\bar{\psi}_Lm\psi_R + \bar{\psi}_Rm^{\dagger}\psi_L
\eeq
is of course {\em not} invariant under $\psi_L \to U_L\psi_L,
\psi_R \to U_R\psi_R$. But if the mass (matrix) were a field 
transforming like $m \to U_LmU_R^{\dagger}$ then the QCD Lagrangian
with this mass term {\em would} be invariant. Then we know what the
corresponding terms must look like in the chiral Lagrangian. The
leading term will be
\beq
{\cal L} ~=~ \Sigma{\rm Tr}[mU + m^{\dagger}U^{\dagger}]
\eeq
and there will of course be higher order terms involving higher
powers of $m$ and derivatives. Such a chiral Lagrangian will
be invariant under the combined transformation
\beq
m \to U_LmU_R^{\dagger}~~,~~~~~~~  U \to U_R U U_L^{\dagger} ~.
\eeq
A Lagrangian constructed in this way will thus explicitly break
chiral symmetries in precisely the same way as the underlying
theory of QCD. 

\subsection{Different counting schemes}

So far we have defined the chiral Lagrangian by the totally
of all terms that transform properly under chiral transformations.
This looks rather hopeless, since there are infinitely many terms.
Does such a theory have any predictive power at all? Weinberg
\cite{Weinberg} showed that it indeed does. First, since all
possible terms are already included in the Lagrangian, it
is by construction {\em renormalizable} since quantum corrections
cannot produce terms not already present in the tree-level
Lagrangian. This goes counter to everybody's intuition about
renormalizable field theories, but it is correct. The chiral
Lagrangian is an {\em effective field theory} and as such
perfectly renormalizable. The theory, however, is not well-defined
at all energy scales, and this is what we should consider next. 
The trouble is the appearance of higher and higher derivatives
in the Lagrangian. Normally, this indicates that theory will 
not be a truly local theory and this may hold here as well. Certainly,
the more derivatives, the more there will be contributions from
the high energy scale. But the theory is not even supposed to
hold at the high energy scale -- it is precisely a low-energy
effective theory. The way to deal with this was explained by
Weinberg: consider an ordering of the Lagrangian where one
expands in the number of derivatives. If this
is to make sense, the Lagrangian must necessarily be viewed as a
theory with a cut-off. In QCD, one would expect the cut-off 
$\Lambda$ to be around
1 GeV; in detail this shows up in the form of the combination
$4\pi F$, which indeed is of that order. Dependence on
this cut-off can be removed for observables below this energy scale. 

The expansion in terms of the number of derivatives can be
made precise, see $e.g.$ ref. \cite{GL}. A counting is introduced
which makes it systematic. Since it is in terms of
derivatives (or momenta), this is referred to as the $p$-expansion.
The pion mass will also enter explicitly in the chiral Lagrangian,
due to non-vanishing quark masses that explicitly break chiral
symmetry in QCD. So an ordering with respect to $m_{\pi}^2$ needs
to be introduced as well. The natural expectation works: one can
systematicall treat $m_{\pi}$ as being of order momentum $p$. Then
chiral perturbation has only one expansion parameter, $p$. The scale
is given by the expected ultraviolet cut-off  $4\pi F$ so
that the dimensionless expansion parameter is $\sim p/(4\pi F)$.
This expansion works quite well phenomenologically, at least in the
sector of the two lightest quarks (and we shall only be concerned
with this light quark sector here).

The $p$-expansion breaks down eventually if one takes the massless
limit at finite four-volume $V$. This is most easily seen by
considering the pion propagator
\beq
\Delta(p^2) ~=~ \frac{1}{V}\frac{1}{p^2 + m_{\pi}^2} ~.
\eeq 
For simplicity, let us take the finite volume $V$ to be a symmetric
four-torus of length $L$. The smallest (quantized) momentum is thus
of order $1/L$. In a massless theory this makes the propagator
$\Delta(p^2)$ vanish for large $L$ like $L^{-2}$. But for the 
momentum {\em zero mode} the cut-off is given entirely given by
the mass. Clearly, for $m_{\pi} \gg 1/L$ the propagator is still
protected by the mass, but as $m_{\pi}L \sim 1$ or much smaller we enter a new
regime. From this point and onward the usual perturbative expansion cannot
make sense in the zero-mode sector. Is this of importance? Can we
not just ignore the zero mode? The answer is unfortunately no. Although
it concerns only one mode, this one mode potentially
overwhelms all other contributions
from the perturbative expansion. The resolution of this problem is
to treat the zero-mode sector exactly in a sense that will be more
clear below \cite{Neuberger,GL1}. Gasser and Leutwyler developed
a modified perturbation theory that includes this feature, while
retaining all other properties of the usual pertubative expansion
of chiral Lagrangians. This has become know as an $\epsilon$-expansion
(not to be confused with the expansion of similar name that
expands in dimensionality $4-d$). The non-zero momentum modes still
retain their usual pertubative expansion since the propagator for
these modes will go at least as $L^{-2}$, even in the chiral limit.

A systematic chiral counting for this so-called $\epsilon$-regime
is as follows \cite{GL1}. Let $\epsilon \sim 1/L$. Then
\beq
m_{\pi} \sim \epsilon^2~,~~~~ p \sim \epsilon ~.
\eeq
It is more instructive to think of this in terms of the microscopic
degrees of freedom. Let us restrict ourselves to two degenerate
light quarks of mass $m$. Due to the Gell-Mann--Oakes--Renner relation
\beq
F^2m_{\pi}^2 ~=~ 2m\Sigma 
\eeq
where $\Sigma$ is the chiral condensate, we see that $m \sim \epsilon^4$.
In the $\epsilon$-regime the quark masses scale as inverse powers of
the volume. In fact, we can think of this as a regime where the extreme
chiral limit $m\to 0$ is taken in a way that correlates with the
way the four-volume $V$ is sent to infinity. The relevant proportionality
factor must have dimension 3, and indeed the right way to think of it
is that the combination $m\Sigma V$ is kept of order unity. Here the
condensate $\Sigma$ provides the missing constant of proportionality,
which is directly related to the fact that the combination
$m\bar{\psi}\psi$ is a renormalization group invariant, in fact it
is the explicit chiral symmetry breaking term in the QCD Lagrangian. 

There is another way in which we can understand the appearance of the scale
$m\Sigma V$. In two-flavor QCD, 
the chiral condensate $\langle\bar{\psi}\psi\rangle$ 
at finite quark mass $m$ and finite volume $V$ is
the trace of the quark propagator,
\begin{eqnarray}
\langle\bar{\psi}\psi\rangle & = & \frac{1}{2V}\partial_m\ln Z \cr
& = & \frac{1}{V}\left\langle {\rm Tr}\frac{1}{D + m}\right\rangle \cr
& = & \frac{1}{V}\left\langle \sum_j\frac{1}{i\lambda_j + m}\right\rangle \cr
& = & \frac{1}{V}\left\langle \sum_{j>0}\left(
\frac{1}{i\lambda_j + m} + \frac{1}{-i\lambda_j + m}\right)
\right\rangle  \cr
& = & \frac{1}{V}\left\langle \sum_{j>0}\frac{2m}
{\lambda^2_j + m^2}\right\rangle 
\end{eqnarray}
where we have used that each quark contributes to 
$\langle\bar{\psi}\psi\rangle$ with the same amount. Use has also been
made of the fact that every non-vanishing eigenvalue $\lambda_j$
can be matched by it opposite-sign counterpart. This
follows directly from the spectrum being chiral, $\{D,\gamma^5\} = 0$.
The above spectral representation
can be used to deduce the so-called
Banks-Casher relation for the infinite-volume chiral condensate:
\beq
\Sigma ~=~ \lim_{m\to 0}\lim_{V\to \infty}
2m\int_0^{\infty}d\lambda~\frac{\rho(\lambda)}{\lambda^2+m^2} ~=~ \pi\rho(0)
\label{bankscasher}
\eeq
This is a formal expression that should be understood only in terms
of a cut-off theory. The chiral condensate $\langle\bar{\psi}\psi\rangle$
is ill-defined in the ultraviolet, and indeed one expects the spectral
density $\rho(\lambda)$ to behave like $\sim \lambda^3$ (as in a free
theory) for very large momenta, and this would lead to a quadratic
divergence 
$\langle\bar{\psi}\psi\rangle \sim m\Lambda^2$. A term like
$\sim m \ln(\Lambda)$ is also expected. However, the important point
here is not the behavior at the ultraviolet end of the spectrum, but
near the origin. Whether chiral symmetry is spontaneously broken or
not is determined by whether or not the spectral density 
$\rho(\lambda)$ vanishes there. If $\rho(0)$ is to be non-vanishing,
the discrete eigenvalues $\lambda_j$ must accumulate there
at a rate inversely proportional to the volume \cite{LS}:
\beq
\Delta\lambda ~\sim~ 1/V ~~~~~~~{\rm near}~~ \lambda \sim 0
\eeq
so as to yield a finite density at the origin. The inverse of 
the proportionality factor is precisely the condensate
$\Sigma$ on account of the Banks-Casher relation. We thus find
that $\lambda\Sigma V$ will be of order unity for the smallest
eigenvalues if we are to generate a chiral condensate.
As will become clear shortly, there is an $\epsilon$-regime 
of a {\em partially quenched} chiral Lagrangian hidden in this 
statement. 

Before defining the partially quenched theory, let us return
to the integral representation of $\langle\bar{\psi}\psi\rangle$.
If we know how this condensate as a function of the mass $m$,
can we not invert the relation to get $\rho(\lambda)$? The trouble
is that $\rho(\lambda)$ is also a function of quark masses
$m$. If it were not, we could indeed simply invert the relation.
The trick to achieve this is to do partial quenching: one
introduces new unphysical quarks that are used to produce an
intermediate generating function, but which are removed at the
end. The spectral density will then be independent of the masses
of these physical quarks, but if we know the analogue of their
condensates $\langle\bar{\psi}\psi\rangle$, we can invert
the integral representation and compute the spectral density. There are
two ways of introducing such partially quenched quarks: one is
by means of a graded structure \cite{BG}, the other by means
of replicas \cite{DS,DSreplica}. Both methods work beautifully in 
perturbation theory, and for the same reason: observables computed
in perturbation theory will depend polynomially on the number
of additional fermionic $N_f$ and bosonic $N_b$ species. If both
fermions and bosons are kept, the anticommutation sign from fermions
just cancel the corresponding contribution from the fermions. So
an equal number of degenerate fermions and bosons do not contribute to
the partition function, a statement that looks entirely trivial
if phrased formally in terms of functional determinants. 
In the replica method, one keeps only
the fermions (or, equivalently, only the bosons) and instead
removes them from the partition function by taking the replica
limit $N_f \to 0$. Since the behavior is polynomial in perturbation
theory, it is unambiguous how to do so in that setting.

In the $\epsilon$-regime, integrals have to be done exactly,
beyond perturbation theory. Curiously, then, both the graded
method and the replica method run into technical difficulties.
In the graded method it turns out that the naive extension of
the chiral Lagrangian to a graded coset 
fails to yield convergent integrals, a problem that never surfaces
in perturbation theory. This
problem can be overcome by a judicious choice of bosonic integration
\cite{DOTV}: The graded integration domain must be a combination of
a compact and non-compact regions. 
In the replica framework the difficulty
arises from the precise specification of how to analytically
continue in the integer $N_f$. In perturbation theory this
is trivial (because dependence on $N_f$ is polynomial only), 
but beyond perturbation theory one needs more
structure as guidance -- knowing a function at all integers is of 
course not sufficient for defining an unambiguous analytic continuation
into the real line. Fortunately, the required structure is precisely present
in the cases of interest \cite{SplitV}, in fact there is a remarkable
connection to the theory of integrable systems and the so-called
Toda lattice hierarchy. It is beyond the scope of these simple
lectures to go into that fascinating subject, to which we here
only refer to the literature \cite{SplitV1}. It was based on this
structure, and its relation to the graded analog, that the partially
quenched partition function of the chiral Lagrangian was first
computed in the $\epsilon$-regime \cite{SplitV,Fyodorov}.

\section{What is Random Matrix Theory?}

What is a ``random matrix''? It is perhaps not obvious, but
with the conventional definition we must fix its size. Let us
first, for simplicity, consider square matrices of size $N\times N$.
Elements in this matrix can now be chosen at random, according to
our chosen distribution. But this cannot be the whole story, because
some matrices have particular symmetries and particular
transformation properties under conjugation. Indeed, a
sensible definition of Random Matrix Theory is based on a division
into certain classes of random matrices, a classification that dates
back to Dyson \cite{Dyson}. The three main classes are labelled by
an index $\beta$, which can take values 1,2 or 4. Here we will focus 
the $\beta=2$ class where one picks the random matrices to be Hermitian,
and thus having real eigenvalues.

To produce a Random Matrix Theory one does as in statistical mechanics
and {\em sums over an ensemble} of a given set of random matrices
$M$. This produces a partition function,
\beq
Z ~=~ \int dM~ \exp\left[-N{\rm Tr}(M^2)\right] ~,
\eeq
where, for simplicity, we have here restricted ourselves to a 
Gaussian distribution. We can associate with this a probability
distribution
\beq
P(M) ~=~ \frac{1}{Z} \exp\left[-N{\rm Tr}(M^2)\right]
\eeq
and for Hermitian matrices the measure is
\beq
dM ~=~ \prod_{i=1}^NdM_{ii}
\prod_{i<j} d({\rm Re} M_{ij})d({\rm Im} M_{ij}) ~.
\eeq

We are not interested in the matrices themselves, but rather in their
real eigenvalues $\lambda_i$. It is therefore advantageous to go to
an eigenvalue representation of the partition function by means of
diagonalization. This is achieved by a unitary transformation:
\beq
M ~=~ U^{\dagger}DU~,~~~~ D = {\rm diag}(\lambda_1,\ldots,\lambda_N)
\eeq
with $U^{\dagger}U = 1$. 

We note that the Hermitian matrix $M$ is described by $N^2$
coefficients, while the diagonal matrix $D$ has $N$ coefficients
and $U$ has $N^2-N$. So we can indeed make the change of variables
$$
           M~~\to~~ D,U ~.
$$
Since the Haar measure on $U(N)$ is left and right invariant, 
the asociated Jacobian $J$ of this change of variables can only depend
on $D$, $i.e.$, $J = J(\lambda_1,\ldots,\lambda_N)$ and
\beq
dM ~=~ \left(\prod_{i=1}^Nd\lambda_i\right) 
J(\lambda_1,\ldots,\lambda_N) dU ~.
\eeq
To compute $J$, it suffices to go near the identity in $U(N)$:
\beq
 U ~=~ 1 + i\varepsilon
\eeq
where, for unitarity, $\varepsilon^{\dagger} = \varepsilon$. Then
\beq
M ~=~ U^{\dagger}DU ~=~ D - i[\varepsilon,D]
\eeq
or, in components,
\beq
M_{ij} ~=~ \lambda_i\delta_{ij} + i\varepsilon_{ij}
(\lambda_i - \lambda_j)~~ {\rm [no~ sum~ on~} i,j{\rm ]} ~.
\eeq 
In other words,
\begin{eqnarray}
i = j &:& dM_{ii} = d\lambda_i \cr
i < j &:& d({\rm Re} M_{ij}) = (\lambda_i-\lambda_j)
d({\rm Re}(i\varepsilon_{ij})) \cr
&~& d({\rm Im} M_{ij}) = (\lambda_i-\lambda_j)
d({\rm Im}(i\varepsilon_{ij}))
\end{eqnarray}
so that
\begin{eqnarray}
dM &=& \left(\prod_{i=1}^NdM_{ii}\right)
\prod_{i<j} d({\rm Re} M_{ij})d({\rm Im} M_{ij} \cr
&=& \left(\prod_{i=1}^Nd\lambda_i\right)\prod_{i<j}
(\lambda_i-\lambda_j)^2
\end{eqnarray}
up to the group volume of $U(N)$. Note also that
\beq
{\rm Tr}(M^2) ~=~ {\rm Tr}(U^{\dagger}DUU^{\dagger}DU)
~=~ {\rm Tr}(D^2) ~=~ \sum_{i=1}^N\lambda_i^2
\eeq
so that also the probability measure can written entirely
in terms of eigenvalues and the partition is then
\beq
Z~=~ \int \left(\prod_{i=1}^Nd\lambda_i\right)\prod_{i<j}
(\lambda_i-\lambda_j)^2e^{-N\sum_{i=1}^N\lambda_i^2} ~.
\label{ZGauss}
\eeq
There is an easy mnemonic to see that this Random Matrix Theory 
is in the $\beta=2$ class since it can be read off from the power ``2'' in 
$$
\prod_{i<j}(\lambda_j-\lambda_i)^2 ~.
$$
This particular combination of eigenvalues can
be written in the form of a so-called Vandermonde determinant
raised to the second power:
\beq
\prod_{i<j}(\lambda_j-\lambda_i)^2 ~=~ \Delta(\{\lambda_i\})^2
\eeq
where
\beq
\Delta(\{\lambda_i\}) ~=~ \det(\lambda_j^i) ~~~~~j = 1,\ldots n, ~i = 0
\ldots n-1~ .
\eeq

The Random Matrix Theory (\ref{ZGauss}) is of course well defined
for all $N$. When $N$ is finite, the $N$ eigenvalues will jump
around in a rather random manner. But the presence of the 
Vandermonde determinant (squared) in front of the Gaussian
damping factor has an important consequence: {\em level repulsion}.
The probability of two eigenvalues approaching each other goes to zero.
For finite $N$, there could be two ways to get around this:
1) all eigenvalues could spread all over the real line or 2)
eigenvalues could be confined to a finite interval, but level
repulsion would force their distributions to lock in an essentially
equidistant form. This will clearly depend on how we rescale eigenvalues.
It turns out that the Gaussian damping
factor in the probability distribution (\ref{ZGauss}) is so strong that 
eigenvalues are forced to lie on a finite interval, and it
is thus the 2nd realization of eigenvalue distributions we are seeing
with such measures. The factor of $N$ in the exponent
is precisely inserted so
as to ensure this compact distribution without further rescalings
of the eigenvalues.

The distribution of eigenvalues becomes beautifully simple in
the limit of very large $N$: we get the famous Wigner semicircle
distribution.

\subsection{Chiral Random Matrix Theory}

We are now ready to consider a Random Matrix Theory that is
chiral. The identification of this class of theories to be the
one relevant for low-energy QCD was first made by Verbaarschot
in a remarkable series of papers about 15 years ago \cite{SV,V}.
The idea was to base the Random Matrix Theory on transformation
properties of the Dirac operator $D$ in QCD. In particular, since
$D$ anticommutes with $\gamma^5$, $\{D,\gamma_5\} = 0$,
a random matrix $M$ is introduced that has a similar structure. Moreover,
gauge field configurations can be assigned to different classes
depending on their topology, which is classified according to the invariant
\beq
\nu ~=~ \frac{g^2}{16\pi^2}\int d^4x{\mbox Tr}(\tilde{F}_{\mu\nu} 
F^{\mu\nu})
\eeq 
where $\tilde{F}_{\mu\nu} = (1/2)\epsilon_{\mu\nu\alpha\beta}F^{\mu\nu}$
is the dual QCD field strength tensor.
In the QCD partition function one sums over all these different
classes that have $\nu=0, \pm 1, \pm 2, \pm 3, \ldots$, but it
can be useful to consider also sectors of fixed $\nu$. A
well known mathematical theorem ensures that in such sectors
there are exact zero modes of the Dirac operator, all of definite
chirality. The number of positive minus negative chirality
zero modes is precisely the index $\nu$. Leutwyler and Smilga
\cite{LS} suggested to study properties of the Dirac operator
in such sectors of fixed topology by means of the relationship
to the chiral Lagrangian. This gives an exciting new
opportunity for cross-checking
how low-energy observables in the full theory (QCD) are in one-to-one
correspondence with observables in the effective large-distance
theory of the chiral Lagrangian.

Verbaarschot and Shuryak \cite{SV} introduced the following
chiral Random Matrix Theory:
\beq
Z ~=~ \int dM \left(\det(M)\right)^{N_f} \exp\left[-(N/2){\rm Tr}V(M^2)\right]
\label{chiralRMT}
\eeq
where $V(M^2)$ will be discussed below, and where
$$
M ~\equiv~
\left( {\begin{array}{cc}
 0 & W^{\dagger}  \\
 -W & 0  \\
 \end{array} } \right)
$$
is a $(2N + \nu)\times(2N + \nu)$ block Hermitian matrix composed of
the complex $N \times (n+\nu)$ rectangular matrix $W$. The integral
is over the elements of this matrix. We note that both measure
and integrand is then invariant under
\beq
W \to V^{\dagger}WU~,~~~~~~ U \in U(N)~,~~~~~ V \in U(N+\nu)
\eeq
and for this reason this Random Matrix Theory is called chiral unitary.
But why is it `chiral'? Consider the $(2N + \nu)\times(2N + \nu)$ matrix
$$
\gamma^5 ~\equiv~
\left( {\begin{array}{cc}
 {\mathbf 1}_N & 0  \\
 0 & -{\mathbf 1}_{N+\nu}  \\
 \end{array} } \right)
$$
This matrix clearly anticommutes with the matrix $M$,
\beq
\{M,\gamma^5\} ~=~ 0 ~,
\eeq
which implies that all non-zero eigenvalues of $M$ come paired
$\pm \lambda_j$. So the matrix $M$ shares the $U(1)$ chirality property
of the full Dirac opertator of QCD. Moreover, due to its 
rectangular block decomposition, there are $\nu$ zero modes. The spectrum
of $M$ therefore has two parts: there are $2N$ non-zero eigenvalues
that are chirally paired and $\nu$ zero modes. 

Actually, the chiral Random Matrix Theory (\ref{chiralRMT}) was
originally introduced with just a Gaussian measure, but it was
believed from the beginning that the so-called `microscopic'
results that could be derived from it would be {\em universal},
$i.e.$ to a very large extent independent of the chosen potential.
Indeed, it was proven in ref. \cite{ADMN} that all results to
be discussed below hold for an arbitrary potential
\beq
V(M^2) ~=~ \sum_{k\geq 1}\frac{g_k}{k} M^{2k}
\label{pot}
\eeq
with, essentially, only one single constraint: the spectrum of
eigenvalues of $M$ must have support at the origin: $\rho(0) 
\neq 0$. This is in beautiful accord with the intuition, to be
be made more precise below, that a non-vanishing chRMT spectral
density at the origin is required for this theory to describe
aspects of chiral symmetry breaking -- in heuristic analogy with
the Banks-Casher relation in QCD. The universality and robustness
of this result is crucial for the understanding 
of why chiral Random Matrix Theory
can describe chiral symmetry breaking in a full-fledged quantum
field theory such as QCD.

Interestingly, one can also consider a chiral Random Matrix Theory
where the spectral density at the origin $\rho(0)$ is carefully
tuned so as to produce a zero: $\rho(0) = 0$ \cite{ADMN1}. One
might hope that this could represent chiral symmetry restoration,
one way or another. This, however, has never been established.
There are still universality classes in this case, one for each order
of zero of the spectral density. Each class can be reached, successively,
by tuning more and more of the parameters $g_k$ in (\ref{pot}). This
is an example of universality in what can be dubbed multicriticality
(while the more general condition $\rho(0) \neq$ corresponds to
the non-critical case). A pity that there seems to be no known
instances of quantum field theory where these more sophisticated
universality classes can be of relevance! In any case, the
mere existence of these extended classes shows that
the universality domain of the non-critical ensemble is restricted
and by no means trivially implied for any chiral Random Matrix
Theory potential $V(M^2)$.

Very similar to the way we could introduce an eigenvalue
representation (\ref{ZGauss}) for the non-chiral (Gaussian) ensemble,
one can also find an eigenvalue representation
\begin{eqnarray}
Z &=& \int_{-\infty}^{\infty}\prod_{i=1}^N\left(
dz_i^2~ z_i^{2(N_f+\nu)}e^{-NV(z_i^2)}\right) \left(
\det_{ij}z_j^{2(i-1)}\right)^2 \cr
&=& \int_{0}^{\infty}\prod_{i=1}^N\left(
d\lambda_i~ \lambda_i^{N_f+\nu}e^{-NV(\lambda_i)}\right) \left(
\det_{ij}\lambda_j^{i-1}\right)^2
\label{chev}
\end{eqnarray}
where the $z_i^2$'s are the real (and positive) non-vanishing
eigenvalues of $W^{\dagger}W$, and where in the last line we
have just transformed to more convenient variables $\lambda_i$.
We have also used the fact that diagonalization of $W^{\dagger}W$
automatically gurantees factorization of the potential part in
the manner shown.
It is interesting to note how the dependence on $N_f$ and $\nu$
has merged into the combination $N_f+\nu$ only. This is a
simplifying property that holds only in the massless case. The
case of massive quarks will be discussed below. The square
of the Vandermonde determinant has appeared just as in
the non-chiral case. Also here it will result in an eigenvalue
repulsion.

There are standard methods for computing the spectral correlation
functions of a Random Matrix Theory with an eigenvalue
representation such as (\ref{chev}). A convenient method
is based on orthogonal polynomials, using a beautiful
formalism that dates back to Dyson. Limitations of space
does not permit a detailed discussion of this, and here we
will restrict ourselves to explaining the main results that
follow from that kind of analysis.

Let us from now on work entirely in rescaled variables
$\zeta \equiv 2\pi N\rho(0)\lambda$. For fixed $\zeta$ this implies
smaller and smaller eigenvalues -- those near the origin.
A central object is the {\em kernel} $K(\zeta_1,\zeta_2)$. If we know
this kernel we know all spectral correlation functions
\beq
\rho(\zeta_1,\ldots,\zeta_k) = \det_{ij}K(\zeta_i,\zeta_j)
\eeq
and in particular the spectral density itself,
\beq
\rho(\zeta) ~=~ K(\zeta,\zeta) ~.
\eeq
The kernel appropriate for QCD is the {\em Bessel kernel} derived
by Verbaarschot and Zahed \cite{V},
\beq
K(\zeta_1,\zeta_2) = \sqrt{\zeta_1\zeta_2}\frac{\zeta_1J_{\alpha+1}(\zeta_1)
J_{\alpha}(\zeta_2) - \zeta_2J_{\alpha}(\zeta_1)J_{\alpha+1}(\zeta_2)}
{\zeta_1^2-\zeta_2^2} ~,
\eeq
where $\alpha = N_f+\nu$ (we consider only positive $\nu$ here).
Despite appearance, this is regular at coincident points where it
gives the microscopic spectral density
\beq
\rho(\zeta) = \frac{\zeta}{2}\left(J_{\alpha}(\zeta)^2 - 
J_{\alpha+1}(\zeta)J_{\alpha-1}(\zeta)\right) ~.
\label{rhomicro}
\eeq 
This is the most simple and remarkable result of the chiral
Random Matrix Theory approach. The claim is that this gives
the microscopic spectral density of QCD with $N_f$ massless
quarks in the finite-volume regime we consider. Below we shall
see why this is so, and how to include massive quarks as well.
Amazingly, if we boldly set $N_f=0$ this should give the 
spectral density of the Dirac operator in pure Yang-Mills theory!
When we see the connection with the partially quenched chiral
Lagrangian this makes it clear why this should be possible
(with some caveats). Another immediate check on the above spectral
density comes from the fact that it reproduces exact spectral
sum rules derived by Leutwyler and Smilga \cite{LS} on the basis of the
chiral Lagrangian alone.

\subsection{Spectral properties of the Dirac operator}

The pioneering paper that showed how these detailed predictions for
the spectral density of the Dirac operator can be tested in
lattice gauge theory simulation was ref. \cite{Tilo}. Quenched staggered
fermions with gauge group $SU(2)$ was used there. Because the
chiral symmetries of staggered fermions away from the continuum
limit are different from those of continuum fermions
when the fermions carry real or pseudo-real
representations of the gauge group (pesudo-real in the case of
fundamental fermions and gauge group $SU(2)$), the detailed
predictions were not those of eq. (\ref{rhomicro}). But also for this
case the analytical prediction had been worked out \cite{Nagao} and
although it differs in detail, it shares many of the properties
with that of (\ref{rhomicro}).
The analytical expressions for continuum fermions based on this
gauge group and fermions in the fundamental representation had
been worked out as well \cite{Vsu2}. We shall not go into great
detail with these different predictions, but it is an amazing
fact, first recognized by Verbaarschot \cite{Vthree}, that there
is a one-to-one correspondence between the Dyson classification
of (chiral) Random Matrix Theory and the three chiral symmetry
breaking patterns in vector-like gauge theories:
\begin{itemize}
\item The fermion representation $r$ is pseudo-real: Chiral 
symmetries are enhanced from $SU(N_f)\times SU(N_f)$ to 
$SU(2N_f)$, and the expected symmetry breaking pattern is
$SU(2N_f) \to Sp(2N_f)$.
\item The fermion representation $r$ is complex: The expected symmetry
breaking pattern is $SU(N_f)\times SU(N_f) \to SU(N_f)$.
\item The fermion representation $r$ is real: Chiral symmetries are again 
enhanced to $SU(2N_f)$, and the expected chiral symmetry breaking pattern
is $SU(2N_f) \to SO(2N_f)$.
\end{itemize}
That there is indeed a one-to-one matching with the three Dyson
classes, and that the chiral symmetry breaking patterns really
do follow the above classification, has been tested on the lattice
in great detail for a variety of different fermion representations
and it works in all cases \cite{reps}.

There is one further general peculiarity of simulations with staggered
fermions on the lattice which in the beginning caused some confusion,
but which really is just one further confirmation of the above
classification scheme. This arises from the fact that staggered
fermions away from the continuum limit have an additional $U(1)$
factor in the coset of chiral symmetry breaking (this holds for all
three classes). As we will show below, this means that all results
for staggered fermions away from the continuum are {\em equivalent}
to those we would get if we had projected on the $\nu = 0$ sector.
Any attempt at a further projection on sectors with $\nu \neq 0$
leads to the same partition function. So for staggered fermions
away from the continuum it will {\em look} as if only the sector
of $\nu=0$ appears. This has been confirmed by a high-statistics
study where gauge field configurations were even selected according
the their (approximative) topological indices: all spectral data
of the staggered Dirac operator look as if they have been obtained
from the $\nu = 0$ sector \cite{Kari}. This should of course change
as the continuum limit is reached. By using highly improved
actions as well as improved Dirac operators that smear both 
configurations and observables it has been checked that predictions
of continuum fermions do match those of staggered fermions 
eventually \cite{Hart}.

While most initial studies of the spectral properties of the Dirac
operator were restricted to gauge group $SU(2)$, results were
soon available for the QCD gauge group of $SU(3)$ \cite{su3},
nicely confirming the prediction (\ref{rhomicro}). A far more exciting
possibility was to check all of these predictions with `chirally
good' fermions based on the overlap operator. This was first done
in ref. \cite{Edwards}. Here the effect of topology can be 
clearly seen, and this works well even for approximate realizations
of the overlap operator \cite{Lang}. A high-statistics study with several
eigenvalues included has been made in ref. \cite{Luscher}, showing clear
agreement with the predictions. It is also interesting to see
how, with an infinite number of colors, chiral symmetry can break
even at finite four-volume. Results still fall right on
the chiral Random Matrix Theory predictions, now as a result of
scaling in the number of colors $N_c$ \cite{Rajamani}. Here one sees
clearly how color and volume acquire complementary roles at
large-$N_c$, the corresponding scaling variable being
$\lambda\Sigma N_c V$.

So far our discussion has been restricted to the massless limit,
which is easily taken in lattice gauge theory if one restrict
oneself to the quenched theory. An interesting question is how to 
include quark masses in all of these preditions. If we consider
the chiral Random Matrix Theory (\ref{chiralRMT}), it is clear
that masses $m$ must be rescaled at the same rate as eigenvalues
$\lambda_i$. Let us re-emphasize that $m$ in (\ref{chiralRMT})
is simply a dimensionless parameter in the matrix integral. It
is {\em not} a quark mass. However, just as the universal predictions
for rescaled eigenvalues $N\rho(0)\lambda$ correspond to
volume-rescaled eigenvalues $V\Sigma\lambda$ on the QCD side,
so $N\rho(0)m$ will correspond to the scaling variable
$V\Sigma m$ in QCD. The precise predictions were worked out
in ref. \cite{DN}. The clue to the solution was there found in
the theory of orthogonal polynomials, which is known to solve
the spectral problem of eigenvalue representations like those
of (\ref{chiralRMT}). Intuitively, it comes about as follows:
if we know a set of orthogonal polynomials corresponding
to (\ref{chiralRMT}) {\em without} the determinant in the
measure, we can construct the set of otrhogonal polynomials
with {\em one determinant} in the measure. Basically, one
expands the new set of orthogonal polynomials in the basis of
the old set. The coefficient in this expansion can be fixed
by what is known as Christoffel's Theorem. This solves the
$N_f=1$ case in terms of the quenched case. To illustrate,
one finds the spectral density in sector of topological \cite{DN}
charge $\nu=0$,
\beq
\rho(\zeta,\mu) = \frac{\zeta}{2}\left(J_0(\zeta)^2 + J_1(\zeta)^2\right)
-\zeta\frac{J_0(\zeta)[I_1(\mu)\mu J_0(\zeta)+I_0(\mu)\zeta J_1(\zeta)]}
{(\zeta^2+\mu^2)I_0(\mu)}
\eeq
In this way of writing it, we nicely illustrate an essential
feature:
\beq
\lim_{\mu\to\infty} \rho(\zeta,\mu) = \rho(\zeta)
\eeq
in other words: when we take the (rescaled) quark mass to infinity 
we recover the quenched spectral density. This is as we would intuitively
expect, since the quark should then decouple and the eigenvalues
should distribute themselves according to the quenched distributions.

All of this was generalized in ref. \cite{DN} to any number of
flavors, and the universality proof was there extended to this generalized 
case as well. Moreover, the explicit expressions for the massive kernels
were provided there. From these, all spectral correlation functions
with massive quarks follow. However, the expressions can become
unwieldy, and there is actually a much more compact way of
understanding these results. It turns out that, quite surprisingly,
one can expression all microscopic spectral correlators compactly
in terms of finite volume QCD effective partition functions.
The crucial ingredient is a powerful identity derived by Zinn-Justin
\cite{Zinn}
in the case of the ordinary unitary ensemble, and which is readily
generalized to the chiral case. Let us consider the $\nu=0$ case
in detail \cite{me}. The kernel can then be expressed as
\begin{eqnarray}
K_N(z,z';m_1,\ldots,m_{N_{f}})\! &=& \!
\frac{e^{-\frac{N}{2}(V(z^2)+V(z'^2))}\sqrt{zz'}
\prod_{f}\sqrt{(z^2+m_f^2)(z'^2+m_f^2)}}{
\tilde{\cal Z}_{\nu}^{(N_{f})}(m_1,\ldots,m_{N_{f}})} ~\times \cr
&& \!\!\!
\int_0^{\infty}\! \prod_{i=1}^{N-1}\!\!\!\left(\!d\lambda_i 
\lambda_i^{\nu}(\lambda_i\!-\!z^2)(\lambda_i\!-\!z'^2)\!\!
\prod_{f=1}^{N_{f}}(\lambda_i\! +\! m_f^2)
{\rm e}^{-NV(\lambda_i)}\!\right)\!\!\left|{\det}
\lambda_j^{i-1}\right|^2 ~.
\end{eqnarray}
The last integral is over $(N\!-\!1)$ eigenvalues only. However, in the
large-$N$ limit we shall consider below, this distinction can be ignored.
Thus, in the large-$N$ limit we have
\begin{eqnarray}
K_N(z,z';m_1,\ldots,m_{N_{f}}) &=&
e^{-\frac{N}{2}(V(z^2)+V(z'^2))}\sqrt{zz'}
\prod_{f}\sqrt{(z^2+m_f^2)(z'^2+m_f^2)}\cr
&& \times \frac{
\tilde{\cal Z}_{\nu}^{(N_{f}+2)}(m_1,\ldots,m_{N_{f}},iz,iz')}{
\tilde{\cal Z}_{\nu}^{(N_{f})}(m_1,\ldots,m_{N_{f}})} ~,
\end{eqnarray}
where the matrix model partition function in the numerator is evaluated
for a theory corresponding to $(N_f\!+\!2)$ fermions, of which two 
have imaginary mass. By means of the usual factorization property, all
higher $n$-point spectral correlation functions are then also explicitly
expressed in terms of the two matrix model partition functions
$\tilde{\cal Z}_{\nu}^{(N_{f})}$ and $\tilde{\cal Z}_{\nu}^{(N_{f}+2)}$.
The spectral density corresponds to the two additional (imaginary)
masses being equal:
\beq
\rho^{(N_{f})}(z;m_1,\ldots,m_{N_{f}}) ~=~
\lim_{N\to\infty} K_N(z,z;m_1,\ldots,m_{N_{f}}) ~.
\eeq
All of this is exact, and at the level of unrescaled variables.
We can now take the microscopic scaling limit in which $\zeta \equiv
z N2\pi\rho(0)$ and $\mu_i \equiv m_i N2\pi\rho(0)$ are kept fixed as
$N\!\to\!\infty$. In this limit the pre-factor 
$\exp[-(N/2)(V(z^2)+V(z'^2))]$ becomes replaced by unity. Identifying
$\Sigma = 2\pi\rho(0)$, this is the limit in which we can compare
with the finite-volume partition function of QCD. Proceeding in this
way, we get, with $C$ being an overall constant,
\beq
\rho^{(N_{f},\nu)}(\zeta;\mu_1,\ldots,\mu_{N_{f}}) ~=~
C |\zeta| \prod_{f}(\zeta^2+\mu_f^2)~\frac{
{\cal Z}_{\nu}^{(N_{f}+2)}(\mu_1,\ldots,\mu_{N_{f}},i\zeta,i\zeta)}{
{\cal Z}_{\nu}^{(N_{f})}(\mu_1,\ldots,\mu_{N_{f}})} ~,
\eeq
and all $n$-point correlation functions are given by one magical
Master Formula \cite{me}
\beq
\rho^{(N_{f},\nu)}(\zeta_1,\ldots,\zeta_n;\mu_1,\ldots,\mu_{N_{f}}) ~=~
\det_{a,b} K(\zeta_a,\zeta_b;\mu_1,\ldots,\mu_{N_{f}}) ~.
\label{correl}
\eeq
This generalizes to any $\nu$ \cite{AD} and in fact leads to
some surprising identities among the partition functions involved,
identities that express the connection to an associated integrable
hierarchy. The advantage is that the finite-volume partition
functions are known in exact analytical forms \cite{Andy}. Results
have also been generalized to the other chiral symmetry breaking
classes \cite{Kanzieper,NagaoN}.

One immediate check on these massive spectral densities is that
they reproduce exact {\em massive} spectral sum rules \cite{Dsum}
just as the massless spectral densities reproduce the massless
spectral sum rules of Leutwyler and Smilga.

Amazingly, the same procedure generalizes to the computation of 
{\em individual eigenvalue distributions}. Such a notion makes clear sense
in the chiral Random Matrix Theory framework since there is a natural
starting point from where to start counting eigenvalues (the origin).
After the lattice gauge theory community had accepted that chiral
Random Matrix Theory gives exact results for the spectral density of
the Dirac operator, and all higher-point spectral correlation functions,
there remained a curious belief that even the very notion of ordered
Dirac operator eigenvalues and their precise distributions could not
be given field theoretic meaning. This is of course not so: ordered
Dirac operator eigenvalues have precisely as much meaning in 
the field theory framework as the spectral density and all spectral
correlation functions. This will become clear below. But as first step
towards this, we should describe how individual eigenvalue
distributions of the chiral Random Matrix Theory can be given 
explicitly in terms of finite volume partition functions for first
the smallest Dirac operator eigenvalue \cite{NDW} and then the
general distribution for the $k$th Dirac operator eigenvalue
\cite{DN1}, as counted from the origin. Let us denote the joint
probability distribution of the smallest $k$ ordered eigenvalues
by $\omega_k(\zeta_1,\ldots,\zeta_{k-1},\zeta_k)$, where
$\zeta_1 \leq \ldots \leq \zeta_k$.
By manipulations very similar to the ones used above, one can show
that it can be written in terms of finite volume
partition functions \cite{DN1}
\begin{eqnarray}
\omega_k(\zeta_1,\ldots,\zeta_{k-1},\zeta_k) &=& C e^{-\zeta_k^2/4}
\zeta_k\prod_{i=1}^{k-1}\left(\zeta_i^{2\nu+1}\prod_{j=1}^{N_f}
(\zeta_i^2+\mu_j^2)\right)\prod_{i>j}^{k-1}(\zeta_i^2-\zeta_j^2)^2
\prod_{j=1}^{N_f}\mu_j^\nu   \cr
&&\times\frac{Z_2(\{\sqrt{\mu_i^2+\zeta_k^2}\},\{\sqrt{\zeta_k^2-\zeta_i^2}\},
\{\zeta_k\})}{Z_{\nu}(\{\mu_i\})}
\end{eqnarray}
where the ``masses'' $\sqrt{\zeta_k^2-\zeta_i^2}$ are doubly degenerate
(and $1 \leq i \leq k-1$). Similarly, the set of ``masses'' 
$\{\zeta_k\}$ is $\nu$-fold degenerate.

A very high-accurate Monte Carlo study of these individual eigenvalue
distributions has been done in ref. \cite{evMC}, for both the quenched
case and for theories with dynamical finite-mass fermions. The agreement
is quite spectacular, with parameter-free fits that lie right on the
analytical predictions. Of course, if one for some reason would fear
these predictions, one could also just look at the chiral condensate
which follows a very precise scaling law based on these analytical
predictions for the chiral condensate \cite{Vcond,wecond}.  

One important question is how individual eigenvalue distributions,
ordered in the way we have discussed above, can be understood and
derived from
the quantum field theory point of view. How do we even define such
distributions in quantum field theory? It turns out that there
is a simple answer \cite{AD1}. For simplicity, let us here restrict
ourselves to the case of just the smallest Dirac operator eigenvalue,
the distributions of the larger eigenvalues follow analogously
\cite{AD1}. Here one can show that the so-called gap probability
has a convergent expansion of the form
\beq
E_0(s) = 1 - \int_0^s \!d\lambda_1~\rho(\lambda_1)
+ \frac{1}{2}\int_0^s\!d\lambda_1d\lambda_2~\rho_2(\lambda_1,\lambda_2)
+ \ldots
\eeq
involving higher and higher spectral correlation functions. All of
these spectral correlation functions can be computed in quantum field
theory by means of suitable sources. The distribution of the smallest
eigenvaluie now follows by differentiation:
\beq
p_1(s) ~=~ -\frac{\partial}{\partial s}E_0(s) ~.
\eeq
For the $k$th eigenvalue this follows by a simple iterative scheme
that involves increasingly higher order correlation function as
the first term in the expansion \cite{AD1}. But the convergence is
fast in the appropriate region of $\lambda$, and this scheme is
in fact a quite practical way of computing these distributions if
one is only interested in high numerical accuracy.

Another issue concerns gauge field topology. It is clear from the
chiral Random Matrix Theory construction that topology is intimately
tied together with whole construction. While a summation over
topology is straightforward in QCD, there seems to be no way to
peform such a summation in the context of chiral Random Matrix
Theory. Every sector corresponds to matrices of different sizes.
However, all physical observables can be decomposed into their
counterparts in distinct topological sectors, and a summation
over topology is therefore indirectly possible through \cite{topo} 
\beq
\langle {\cal O}\rangle ~=~ \frac{1}{Z}\sum_{\nu}
Z_{\nu}\langle {\cal O}\rangle_{\nu}
\eeq
where $Z$ is the full partition function and subscript $\nu$
indicates when the average is taken in a sector of fixed 
topological charge. Note how averages are weighted by the
partition function in fixed sectors. Using this relationship and
the Master Formula derived above, it is possible to perform
the sum analytically and derive an analogous formula for the
spectral density of the full theory \cite{topo}. It is also
particularly illuminating to see analytically how the triviality
of results in the $N_f=1$ theory are recovered from the highly
non-trivial predictions in sectors of fixed topological charge.

\subsection{Matching chiral Random Matrix Theory with the chiral Lagrangian}

The Master Formula (\ref{correl}) gives a strong indication that
all these chiral Random Matrix Theory results can be derived
entirely from chiral Lagrangian framework. This is indeed the
case \cite{James,DOTV}. The framework is partially quenched
chiral perturbation theory, which we have already discussed above.
Through it, one can derive a one-to-one correspondence between
the spectrum derived from the chiral Lagrangian and the spectrum
derived from chiral Random matrix Theory. The final steps showing
that this holds for all spectral correlation functions were taken
by Akemann and Basile in ref. \cite{Basile}.

What has been achieved is thus a precise and detailed understanding
of how chiral Random Matrix Theory can yield exact results. It
provides an exact representation of the leading-order chiral
Lagrangian in the extreme finite-volume $\epsilon$-regime. The
proof goes through the partially quenched theory which here
serves two purposes at one time: (1) it provides predictions
for genuine partially quenched lattice gauge theories in this
finite-volume regime, and (2) it serves as a generating function
of spectral correlation functions. In chiral Random Matrix Theory
one does often not make use of a partially quenched (graded)
version of the theory because other simpler technqiues (such as the
one based on orthogonal polynomials) are available. It is precisely
by going through the graded version that the full equivalence has
finally been established for all correlation functions.

An interesting question is whether one can extract information about
the other low-energy constants $F$ as well as all higher-order
coupling constants of the chiral Lagrangian by means of spectral
properties of the Dirac operator. In principle, this should be
possible since the influence of these couplings on the distributions
of the Dirac operator eigenvalues can computed. By a careful
comparison to numerical lattice data, one should be able to
extract these other low-energy parameters of QCD by considering
only a few of the lowest Dirac operator eigenvalue -- a remarkable
thought. However, in practice it is not that easy. The simplest
example is the leading correction to $\Sigma$, which we already
discussed above. In principle, by considering a variety of 
different volumes $V$, and perhaps different geometries as well,
one can extract the pion decay constant $F$ from this.

Other chiral Random Matrix Theory approaches, aimed specifically
at determining $F$, have also been invented. One consists in
introducing a new source in the chiral Lagrangian that couples
directly to $F$ to leading order in the chiral counting, while
retaining anti-Hermiticity of the Dirac operator in that
context. Although it is to be considered only as a source,
it actually has a physical interpretation as imaginary isospin
potential. With an external isospin potential the QCD measure
is free of the sign problem that plagues QCD with baryon chemical
potential. Making the potential purely imaginary ensures that
the Dirac operator eigenvalues with this type of chemical potential remain
anti-Hermitian. This approach was first developed in the
chiral Lagrangian framework \cite{Fpi}, but inspired by a
construction due to Osborn \cite{Osbornpot} 
who introduced a two-matrix chiral theory do describe baryon 
chemical potential, a chiral two-matrix theory was set up
to describe imaginary isospin chemical potential \cite{ADOS}.
Indeed, results from this theory, which we will describe next,
agree precisely with those obtained from the chiral Lagrangian
\cite{Fpi}. Because all the powerful machinery of the more
usual chiral Random Matrix Theory is preserved, this formulation
allows for an explicit calculation of all spectral correlation 
function.

The first step consists in considering the eigenvalues of 
two different Dirac operators in QCD,
\begin{eqnarray}
D_1\psi_1^{(n)} &\equiv & [D(A)+i\mu_1\gamma_0]\psi_1^{(n)} ~=~
 i\lambda_1^{(n)}\psi_1^{(n)}  \cr
D_2\psi_2^{(n)} &\equiv & [D(A)+i\mu_2\gamma_0]\psi_2^{(n)} ~=~
 i\lambda_2^{(n)}\psi_2^{(n)} ~.
\label{Dmudef}
\end{eqnarray}
Imaginary isospin potential corresponds to $\mu \equiv \mu_1 = -\mu_2$.
Such a source couples directly to $F$ is the chiral Lagrangian
for which the leading-order terms in a suitably defined $\epsilon$-regime
read:
\beq
Z_{\nu}^{(N_f)} =
\int_{U(N_f)} dU \,(\det U)^{\nu}e^{\frac{1}{4}VF_\pi^2
{\rm Tr} [U,B][U^\dagger,B] + \frac{1}{2} \Sigma V
{\rm Tr}({\cal M}^\dagger U + {\cal M}U^\dagger)} ~.
\label{ZXPT}
\eeq
where the matrix $B=$diag$(\{\mu_1\},\{\mu_2\})$ contains
the two chemical potentials introduced above, and ${\cal M}=
\mbox{diag}(m_1,\ldots,m_{N_f})$ is the mass matrix. 
Chemical potential $\mu_1$ has been assigned to $N_1$ of the quarks 
and chemical potential $\mu_2$ to the remaining $N_f-N_1 = N_2$ quarks.

A two-matrix chiral theory that corresponds to this is
\begin{eqnarray}
Z_{\nu}^{(N_f)}& = &
 \int d\Phi  d\Psi~ 
e^{-{N}{\rm Tr}\left(\Phi^{\dagger}
\Phi + \Psi^{\dagger}\Psi\right)}
\prod_{f1=1}^{N_1} \det[{\cal D}_1 + m_{f1}] 
\prod_{f2=1}^{N_2} \det[{\cal D}_2 + m_{f2}] 
\label{ZNf}
\end{eqnarray}
where ${\cal D}$ is defined by
$$
{\mathcal D}_f = \left( \begin{array}{cc}
0 & i \Phi + i \mu_f \Psi \\
i \Phi^{\dagger} + i \mu_f \Psi^{\dagger} & 0
\end{array} \right) \ \ ,\ \ f=1,2\ ~.
$$
The matrices $\Phi$ and $\Psi$ are complex rectangular 
matrices of size $N\times (N+\nu)$ in the same manner as in the
usual chiral ensemble.

Without going into the technical details, an 
eigenvalue representation can also be found for this theory. Let us
define
\begin{eqnarray}
c_1 &=& (1+\mu_2^2)/\delta^2  \ ,\ \ \ \ c_2 \ =\ (1+\mu_1^2)/\delta^2 \cr
d &=& (1+\mu_1\mu_2)/\delta^2 \ ,\ \  1-\tau \ =\ d^2/(c_1c_2)\cr
\delta &=& \mu_2 - \mu_1\ ,
\label{cdtdef}
\end{eqnarray}
Then,
\begin{eqnarray}
{\cal Z}_{\nu}^{(N_f)}
&=& \int_0^{\infty} \prod_i^N\left(dx_idy_i (x_iy_i)^{\nu+1}
\prod_{f1=1}^{N_1} (x_i^2+m_{f1}^2)
\prod_{f2=1}^{N_2} (y_i^2+m_{f2}^2) \right) \cr
&\times&\Delta(\{x^2\})\Delta(\{y^2\})\det\left[I_{\nu}(2 d N x_i y_j)
\right] 
e^{-N \sum_i^N c_1 x_i^2 + c_2 y_i^2 } ~. \label{evrep}
\end{eqnarray}
The two sets of eigenvalues $x_i$ and $y_i$ are obtained after
diagonalizing
\begin{eqnarray}
\Phi_1 &\equiv& \Phi + \mu_1 \Psi \cr
\Phi_2 &\equiv& \Phi + \mu_2 \Psi ~,
\end{eqnarray}
and due to this redefinition the original matrices now become
coupled in the exponent. This looks horribly complicated, but it
actually has some recognizable structure. First of all, the
parameter $\delta$ is what measures the strength of the external
source, the imaginary chemical potential. Next, by a stroke of
luck, a formalism of bi-orthogonal polymials had recently been
invented for precisely this kind of problem by Eynard and Mehta
\cite{Eynard}. Analytical miracles occur, and one ends with very
simple analytical expressions for all spectral correlation 
functions. Particularly interesting spectral observables are those 
that {\em vanish} if $\delta = 0$. This leads to a strong signal
with which the pion decay constant can be measured (since $\delta$
couples to $F$ in the chiral Lagrangian). All results derived
from the chiral Lagrangian in \cite{Fpi} are reproduced in this
Random Matrix Theory approach. Moreover, all spectral correlation
functions can now be expressed in closed analytical forms
\cite{ADOS}. Also individual eigenvalue distributions can be computed 
analytically in this two-matrix theory \cite{ADmu}. Detailed
comparisons with lattice gauge theory data, for a variety of different
volumes and including the finite-volume corrections 
\cite{Tom1}, have been made recently \cite{Lehner}.

\subsection{Beyond chiral Random Matrix Theory}

Once one has realized the connection between the leading-order 
chiral Lagrangian in the $\epsilon$-regime and chiral Random Matrix
Theory, it becomes obvious how to combine the two. The most difficult
part of the $\epsilon$-expansion is precisely the non-perturbative
contribution from the momentum zero modes of the pesudo-Goldstone
bosons. We have seen how a partially quenched chiral Lagrangian had
to be understood in this context just in order to derive properties
of the spectral density. This is extremely useful for lattice gauge
simulations since there one is often interested in doing partial
quenching as an {\em approximation} to a real simulation. For example,
one might wish to scan more parameter values with the same
lattice configurations, or one has difficulty simulating with
light enough fermions. Here partial quenching can provide additional
information.

The first issue concerns pion-loop contributions to the 
leading-order chiral Lagrangian. This was investigated in the
original paper by Gasser and Leutwyler \cite{GL1}. Considering
the contribution to the chiral condensate, they found that the
Lagrangian remains form invariant except for a rescaling of
the $\Sigma$-parameter,
\beq
\Sigma ~\to~ \Sigma_{eff} ~=~ \Sigma\left[1-\frac{N_f^2-1}{N_f}
\frac{\beta_1}{\sqrt{V}} + {\cal O}(1/V)\right]
\eeq
for a four-dimensional torus of volume $V$. The coefficient is
one of an infinite sequence of shape coefficients that appear in
the $\epsilon$-expansion \cite{Hasenfratz}. This rescaled
low-energy constant $\Sigma_{eff}$ appears naturally in the
expansion of any observable in the $\epsilon$-regime. We see
that the correction term disappears for large volume as
$1/\sqrt{V}$, which is of order $\epsilon^2$, as expected. The
expansion thus seems to work well. In the fully quenched case
this is not the case. There a new scale, set by the $\eta'$-mass
in physical terms, enters and there is no longer a simple
one-parameter expansion. As usual, when things can go wrong,
they will go wrong. In this case one ends up with a new term
that grows logarithmically with volume -- a ``quenched finite
volume logarithm'' \cite{Vlog}. Although such a correction
term can be kept under control if one only compares two
different volumes (at least to that order), it is just one
more indication of the fundamental difficulty of quenched
theories. In a physical theory with dynamical quarks such
a logarithm still shows up at higher orders, but always
suppressed by additional inverse powers of $L$, and therefore
harmless. 

Hansen \cite{Hansen} was the first to explore the $\epsilon$-regime
beyond leading order for mesonic correlation functions in chiral 
perturbation theory. This was before Leutwyler and Smilga \cite{LS}
had shown the usefulness of working in sectors of fixed topological
charge $\nu$, and a re-analysis of Hansen's results in such sectors
has been done \cite{Pilar}. The generalization of Hansen's results
to the quenched case was also done there. What is interesting about
mesonic correlation functions in the $\epsilon$-regime is their
{\em polynomial} dependence on the euclidean length scale. In
the $p$-regime correlators have the conventional exponential
fall-off, but in the $\epsilon$-regime, where one is essentially
``inside the pion'', the correlation function is almost flat.
One can understand this very simply. Consider the zero-momentum
projection of the standard massive pion propagator,
\beq
\int d^3x~ \Delta(x) ~=~ \frac{\cosh\left(m_{\pi}(T/2-t)\right)}
{2m_{\pi}\sinh(m_{\pi}T/2)} ~,
\eeq
where $T$ is the (arbitrarily chosen) euclidean time extent. Apart
from the overall factor of $1/m_{\pi}^2$, we can Taylor expand
this in powers of $m_{\pi}^2$ to find
\beq
\int d^3x~ \Delta(x) = \frac{1}{Tm_{\pi}^2}
+ \frac{T}{2}\left[\left(\tau-\frac{1}{2}\right)^2-\frac{1}{12}\right]
+ \frac{T^3}{24}\left[\tau^2(\tau-1)^2 - \frac{1}{30}\right]m_{\pi^2}
+ \ldots
\eeq
The finite-order polynomials of this expansion are precisely
what in the $\epsilon$-regime 
replace the ordinarily exponentially decaying correlation functions.
The pole-term at $m_{\pi}=0$ is the contribution from the momentum
zero-mode, which indeed blows up in the simple expansion above (but
which is treated exactly in the $\epsilon$-expansion). This is the
way in which the $\epsilon$-regime expansion of correlation
functions merge smoothly with the standard $p$-regime correlation functions.
Such an approach has also been developed further, see, for
instance, ref. \cite{Pilar1}. 

An interesting question is whether one can combine the $\epsilon$-regime
with the $p$-regime, either partially quenched or not. Similarly,
one can consider the question of so-called `mixed actions' in this
context (partial quenching where the light quark limit is reached
by using configurations generated with different types of fermions,
for example). Ideally, one would like to be able to do only one
calculation which would be valid in a domain that stretches between
the two different counting regimes. This is an
active area of research at the moment \cite{Hide,Pilar2,Hide1} and
results are currently being compared to large-scale simulations
based on overlap fermions \cite{Hide2}.

\section{Conclusions}

There is a fascinating interplay between the chiral Lagrangian and
chiral Random Matrix Theory. Here we have reviewed some of the
exact relations that have been established. The fact that chiral
Random Matrix Theory provides the leading contribution to the
$\epsilon$-expansion of chiral perturbation theory has turned out
to give an analytically very powerful new view on low-energy QCD.

In lattice gauge theory, where most of the results discussed have
very direct applicability, one is dealing with a discretized version
of the continuum field theory. This induces lattice artifacts
that are not physical. In a first approximation such discretization
errors can be neglected, but in a more careful analysis they must
be taken into account. In fact, measuring the quantitative effects
of such lattice artifacts can be done in a systematic way following
what is known as the Symanzik program of effective lattice field
theory. There is then a modified chiral Lagrangian that includes
discretization effects, but phrased in the continuum language.
One can then also ask for the effect of discretization errors 
on the Dirac operator spectrum \cite{Sharpe}.
Remarkably, chiral Random Matrix Theory can be extended to include
such effects as well \cite{DSV,O}. It is particular interesting
to see how the special case of $N_f=1$, where there is no
spontaneous breaking of chiral symmetries, looks in this context.
The full partition function is, to leading order, a trivial
exponential of the free energy since there is a mass gap even
in the chiral limit. But it contains now two terms: one proportional
to the quark mass $m$, and one proportional to the leading 
lattice-spacing artifact $a^2$. These two terms compete trivially
in the full theory for which the leading term in the effective 
partition function simply reads
\beq
Z ~=~ \exp\left[m\Sigma V - 2W_8V a^2\right]
\eeq
where $W_8$ is a new low-energy constant of Wilson fermions in this
curious $N_f=1$ theory (the constant of 2 in front is a convention
inherited from the general-$N_f$ theory where one has a genuine
chiral Lagrangian). If one splits the partition function into
sectors of fixed index $\nu$, there is a non-trivial
interplay between the two terms above
due to the fixing of the index \cite{ADSV1}. It is quite amazing
how much carries over from the case of spontaneous chiral symmetry
breaking once the index $\nu$ has been fixed.
For the more physical case of a larger number of
dynamical quarks also other observables, such
as space-time correlation functions, can be computed for lattice
gauge theory observables in the $\epsilon$-regime with explicit
$a^2$ corections. A variety of different scaling regimes can be considered.
Such a program is presently being carried out for the
case of Wilson fermions \cite{Shindler}. This is one of the 
clearly identifiable directions
in which one can envision future research within this field.

\end{document}